\documentstyle[preprint,aps]{revtex}
\begin{document}
\draft
\preprint{Exciton droplet -- 12/21/94}
\title
{Exciton condensate in semiconductor quantum well structures}
\author{Xuejun Zhu$^1$, P. B. Littlewood$^2$,
        Mark S. Hybertsen$^2$, T. M. Rice$^{2,3}$}
\address{
$^1$Department of Physics \& Astronomy, Rutgers University,
Piscataway, New Jersey 08855
}
\address{
$^2$AT\&T Bell Laboratories, Murray Hill, New Jersey 07974
}
\address{
$^3$ ETH-Honggerberg
CH-8093, Zurich, Switzerland
}
\maketitle

\begin{abstract}

We propose that the exciton condensate may form  in a
well-controlled way in appropriately arranged semiconductor
quantum well structures. The mean-field theory of Keldysh
and Kopaev, exact in both the high density and the low
density limits, is solved numerically to illustrate our
proposal. The electron-hole pairing gap and the excitation
spectrum of the exciton condensate are obtained. The energy
scales of the condensate are substantial at higher densities.
We discuss how such densities could be achieved experimentally
by generating an effective pressure.

\end{abstract}

\newpage

The issue of Bose-Einstein condensation of excitons has
been extensively studied, following the early suggestion
by Keldysh and Kopaev \cite{ehRev,Keldysh,BiEx}. In the intervening
years, it continues to be a subject of considerable experimental
and theoretical interest
\cite{BOcond,JRK,Nozieres,Bilayer,expt,expt1,CuOexpt}.
Until recently, much of the experimental work has been carried out
on indirect semiconductors \cite{ehRev}.
In view of the great advances made in our abilities to
design and manufacture high quality artificial
semiconductor quantum well (QW) structures,
it appears opportune to investigate in some detail
the possible exciton condensate states in direct gap semiconductors.

The crux of the matter lies in obtaining an exciton fluid at
sufficiently high densities and low temperatures to
realize a condensed phase. Recent experiments
on $Cu_2O$ \cite{CuOexpt} and on $GaAs$ quantum wells in high
magnetic fields \cite{expt1} appear quite promising. However,
detailed and unambiguous
interpretation of these experiments has been difficult
either because of the low density obtained in QWs (where it has
been argued that disorder dominates
the photoluminescence spectrum
\cite{expt}) or because of non-equilibrium and/or time-dependence
of the density \cite{CuOexpt}.
Clearly it would be very desirable to be able to produce exciton fluids
at {\it controlled},
and possibly higher, densities. This has two advantages:
the energy scale of the condensate would be larger and the variation
of key properties with density could be systematically
examined. Other fundamental questions
related to the dynamics of condensation, possible lasing action,
superradiance and coupling to a coherent photon field could be studied
experimentally under controlled conditions.
Our proposal is to tailor the QW parameters in double well
electron-hole systems so as to generate an effective pressure on
part of the exciton fluid and thereby achieve the physical conditions
necessary for a controlled formation of exciton condensates.

To this end, we set up and numerically solve
the mean-field theory (MFT)
proposed earlier by Keldysh and Kopaev \cite{Keldysh,JRK,Nozieres},
exact in both the high and low density limits,
to obtain the total energy as a function of arbitrary densities for a variety
of interlayer separation of the double layer QWs.
The same MFT is also used to study the
excitation properties of the proposed exciton droplet, which are
crucial to its stability against small perturbations such as
finite temperature, interface disorder, and complications arising
from the band structure of the underlying semiconductors.
At $T = 0$ the exciton fluid for the idealized
system is an insulating Bose-Einstein condensate
at all densities, {\it i.e.\/}, there exists a gap to all charged
excitations. Hence it has been called an
excitonic insulator.

We begin by examining the
ideal two-dimensional (2D) electron-hole system, characterized by
the following Hamiltonian:
\begin{equation}
H_{eh} = \sum_i {{\vec P_{i,e}^2} \over {2m_e}}
       + \sum_i {{\vec P_{i,h}^2} \over {2m_h}}
       + \sum_{i<j} {{e^2} \over {\varepsilon|\vec r_{i,e} - \vec r_{j,e}|} }
       + \sum_{i<j} {{e^2} \over {\varepsilon|\vec r_{i,h} - \vec r_{j,h}|} }
       - \sum_{i,j} {{e^2} \over {\varepsilon
\sqrt{|\vec r_{i,e} - \vec r_{j,h}|^2+d^2} }},
\label{Hamil}
\end{equation}
where
$\varepsilon$ is the background dielectric constant.
We ignore the electron-hole exchange in $H_{eh}$ \cite{ehRev}.
We assume that the electron and the
hole layers are infinitesimally thin, and that they
are separated by a distance $d$ \cite{intra,footnote}.
Within the mean-field theory described later and for
$s$-wave pairing between an electron and a hole, spin degrees
of freedom play no role
other than changing the $r_s$ by a factor of $\sqrt{2}$ at a given pair
density
$n$ and will be ignored henceforth.
In this work, the unit mass $m$ is taken to be twice
the reduced mass of the electron ($m_e$) and the hole ($m_h$), or
$2/m = 1/m_e + 1/m_h$.
The unit length is $a_{a.u.} = \varepsilon \hbar^2 / m e^2$, and the
energy unit is $Hartree = e^2 / \varepsilon a_{a.u.}$.
$r_s$ is defined as
$\pi r_s^2a_{a.u.}^2 = 1/n$.

Two obvious limits of this system are: $r_s \rightarrow 0$
and $r_s \rightarrow \infty$. In the former high density limit,
the pairing between the electrons and the holes is weak, and
can be described in a BCS-like language where only properties
near the Fermi surfaces are affected \cite{Keldysh,JRK,Nozieres,BCSbook}.
In the latter low density
limit, excitons are formed and they interact weakly \cite{BiEx}.
It has been
realized that even in the latter limit, a BCS-like description is also
appropriate \cite{Keldysh,JRK,Nozieres}.
In fact, it is exact in the limit of $r_s = \infty$,
where we have single isolated excitons.

Such a formulation is also valid for 2D, which we will use
in this work.
The variational ansatz for the ground state wavefunction,
taking into account the pairing effects, is in the BCS form
\cite{Keldysh,JRK,Nozieres,BCSbook}:
\begin{equation}
|\Psi^0> = \prod_{\vec k} [ u_{\vec k} + v_{\vec k} a_{e,\vec k}^{\dagger}
                                                    a_{h,\vec k}^{\dagger} ]
|vac>,
\label{wfn}
\end{equation}
with $|u_{\vec k}|^2 + |v_{\vec k}|^2 = 1$.
$u_{\vec k}$ and $v_{\vec k}$ are the variational parameters to
be determined for the present system, and $|vac>$ is the vacuum
state in which the valence band is completely filled,
$a_{e,\vec k}^{\dagger}$ creates an electron in the conduction band,
and
$a_{h,\vec k}^{\dagger}$ creates a hole in the valence band.
As in BCS, this wavefunction does not preserve the
particle number. The solutions are obtained most easily by
going to the second quantized form and minimizing
the free energy at a given chemical potential $\mu$,
$f  = <h_{eh}> - \mu <n>$,
where lower cases are used to denote that the quantities
are per unit volume. We find:
\begin{equation}
<h_{eh}>= \sum_{\vec k} \epsilon_{\vec k} v_{\vec k}^2
-\sum_{\vec {k_1},\vec {k_2}} \biggl(
{1\over 2}V^{ee}_{\vec {k_1}-\vec {k_2}} v^2_{\vec {k_1}} v^2_{\vec {k_2}} +
{1\over 2}V^{hh}_{\vec {k_1}-\vec {k_2}} v^2_{\vec {k_1}} v^2_{\vec {k_2}} +
V^{eh}_{\vec {k_1}-\vec {k_2}}
u_{\vec {k_1}}v_{\vec {k_1}}u_{\vec {k_2}}v_{\vec {k_2}}
 \biggr),
\label{Eaverage}
\end{equation}
and
$<n> = \sum_{\vec k} v^2_{\vec k}$.
In Eq.~\ref{Eaverage},
$V^{ee}_{\vec q} = V^{hh}_{\vec q} = 2\pi / q$,
$V^{eh}_{\vec q} = 2\pi e^{-qd}/ q$.
$\epsilon_{\vec k} = k^2 / m$, the kinetic energy
of an electron-hole pair, each with momentum $\vec k$.
Notice that for
$d \ne 0$ and in cases where the electrons and the holes are
not separately neutralized
by the respective dopants, ({\it e.g.\/,} when they
are created by optical pumping), the density-dependent Hartree term in the
total energy needs to be included.
At a fixed density,
it increases energy/pair by $2 d / r_s^2 $
and $\mu$ by $4 d /r_s^2$, without affecting
the gap function $\Delta_k$.
Setting $\partial f / \partial v_{\vec k} = 0$ and considering only
$s$-wave pairing in which case all quantities are functions of
the magnitude of $\vec k$, we find in close analogy to the BCS algebra:
\begin{eqnarray}
\xi_k &=& \epsilon_k - \mu - \sum_{\vec k'} V^{ee}_{\vec k - \vec k'}
(1 - \xi_{k'} / E_{k'}), \\
\Delta_k &=& \sum_{\vec k'} V^{eh}_{\vec k - \vec k'}
\Delta_{k'} / E_{k'},\\
E_k^2 &=& \xi^2_k + \Delta^2_k.
\end{eqnarray}
$E_k$ is identified
as the pair excitation spectrum \cite{factor2}.

Eqs.~4-6 are coupled equations that must be solved self-consistently.
Given a chemical potential $\mu$, we wish to find the corresponding
density parameter $r_s = \sqrt{1/(\pi n)}$
and the energy per pair $E^{pair} = <h_{eh}>/<n>$.
Letting $k = tan(\beta)$, we set up a Gaussian-Quadrature
grid for $\beta$ and convert the
above equations into a matrix form which can be
solved iteratively
\cite{ZLHR}.
Only the angular averages of the potentials $V^{ee}$
and $V^{eh}$ enter and the logarithmic singularities in
the averaged quantities are treated separately \cite{ZLHR}.
The total energy per unit volume is:
\begin{equation}
<h_{eh}> = {1 \over 2} \sum_{\vec k}
\biggl[
\bigl(\epsilon_k + \mu + \xi_k\bigr) {1-\xi_k/E_k \over 2}-
\Delta_k {\Delta_k/E_k \over 2}
\biggr].
\label{etotal}
\end{equation}

Shown in Fig.~1 are
$E^{pair}$ (solid lines) and $\mu$
(dashed lines) as a function of $r_s$
from the present MFT calculation, for $d = 0$ and $d = 1.0$.
For large $r_s$, $E^{pair}$ and $\mu$
approach the correct respective single
exciton binding energy, largest at $d = 0$.
For small $r_s$, $E^{pair}$ approaches the Hartree-Fock result,
$2/r_s^2 - 16/(3\pi r_s) + 2d/r_s^2$, where the
terms are the kinetic energy, the exchange energy and the
Hartree energy, respectively.
Indeed, the present MFT approach describes both limits correctly in a
natural way, and is expected to provide a reasonable
interpolation for intermediate densities.

It is interesting to examine the strength of the pairing effects
and the excitation spectrum of the exciton fluid.
At a given layer separation $d$,
the maximum in the gap function $\Delta_k$ (or the minimum in
the excitation energy $E_k$) first
increases with $r_s$, reaches a peak value, and then decreses
with $r_s$. Its $k$-dependence exhibits some interesting
features as the chemical potential $\mu$ crosses the bottom
of the single-particle band \cite{ZLHR}.
As an illustration, we show for $r_s = 2.66$ and $r_s = 5.90$
\cite{spin} at $d = 1.0$
the gap function
$\Delta_k$ and the pair-excitation spectrum $E_k$ as a function
of $k$ in Fig.~2.

First of all, $E_k$ remains significant, in fact $\sim~0.4~Hartree$,
or $\sim~7~ meV$ in $GaAs$ QWs, at both of
these densities. Secondly, we note that the gap function
$\Delta_k$ exhibits a variation
with $k$, instead of being sizable only very near $k_F$ as
in the BCS case. This is a direct result of the extended $k$-range
of the attractive interaction $V^{eh}$.
Lastly, a comparison of the gap function at the two
densities shows some qualitative differences between them.
For the high density $r_s = 2.66$ case,
the minimum pair-excitation energy in $E_k$ is located
very close in $k$ to where the maximum in $\Delta_k$ occurs,
and both are close to, but not identical to $k_F$.
But for the low density $r_s = 5.90$ case, the maximum
of $\Delta_k$ is now at $k$ very close to zero, and the minimum
of $E_k$ is at $k$ smaller than $k_F$
\cite{ZLHR}.
Although we have emphasized the similarities between the present
electron-hole problem and the BCS superconductivity,
there are essential differences even at the level
of MFT \cite{ZLHR}.

Screening effects will reduce both the gap function $\Delta_k$ and
the minimum excitation energy in $E_k$. But we expect
the calculated excitation energies to be qualitatively correct,
since the very presence of the gap will make screening less effective
on the energy scale of the gap itself. The basic gap structure
of the excitation spectrum should certainly survive temperatures
of a few $K$.

In the mean-field theory, Fig.~1 shows that the
interaction between the excitons is repulsive so that only the gas
phase exists in equilibrium in the uniform, double QW
electron-hole system \cite{2Dmin}. Such systems have been
utilized in some recent photoluminescence experiments, and
suggestive results for Bose-Einstein
condensation have been reported, but only in the presence of strong
magnetic field at low densities \cite{expt,expt1}. We now show
that an altered structure may induce an effective pressure
on the exciton fluid and thereby stablize higher densities.

We consider QW structures with a spatially varying
well width, given by $w = w(\vec r)$.
Within the lowest subband approximation,
the effects of $w (\vec r)$ can be described
by an effective in-plane potential $V(\vec r) = V(w(\vec r))$.
If $ d V / d r \ll V / d$, at equilibrium the
exciton distribution throughout the QW $n(\vec r)$ will be given by:
\begin{equation}
\mu(n(\vec r)) + V(\vec r) = \mu,
\end{equation}
where $\mu$ is the chemical potential of the excitons in
the QW, $\mu(n)$ is the chemical potential of a
uniform excitonic insulator at density $n$.
Since the kinetic energy
in the $z$-direction depends rapidly on $w$, changing
$w$ is a very effective way of controlling the local chemical
potential $\mu(n(\vec r))$, thereby controlling the local
density $n(\vec r)$.

This is illustrated
schematically in Fig.~3, where the well-width on one side
of the double QW is greater over a circle of radius $R$ (the ``cavity")
than that of the rest. Only one side needs to be adjusted,
since charge-neutrality demands that the electron and hole
densities be equal.
A bias electric voltage is applied to spatially separate the electrons
and the holes after optical pumping.
Several nanoseconds after pumping, during which the electrons
and the holes in the individual single layers have all recombined,
we are left with a double-layer electron-hole system.

Notice that for large $r_s$, the chemical potential $\mu(n)$ remains very close
to the single exciton binding energy $|E_{ex}|$ for a wide
range of density (see Fig.~1).
Therefore, if the pumping intensity is
not too great, the chemical potential of the excitons will remain
at $\mu$ slightly larger (smaller in magnitude) than
$-|E_{ex}|$.
At equilibrium, the density inside the
cavity $n_{cav}$ will then be well
approximated by $\mu(n_{cav}) = - |E_{ex}| + |\delta V |$.
For $GaAs/AlGaAs$ QWs, $m_h = 0.4$, $m_e = 0.067$,
$\varepsilon = 13$, we find the present atomic units have length
$a_{a.u.} = 59.8~\AA$, and energy $Hartree = 18.5~meV$.
As an example, we consider the case where the hole layer
width is adjusted.
In order to change the local chemical potential inside the
cavity from $- |E_{ex}|$ to $\mu(r_s = 3.69)$ at $d = 1$ (a value
between the two $r_s$'s considered earlier),
the well-widths can be arranged to be:
$w(\vec r) = 60~\AA$ for $ r > R$ and,
$w(\vec r) = 73~\AA$ for $ r < R$.
We note that, quite generally, any QW structure
similar to that of Fig.~3 will serve to form an exciton
droplet at a density largely controlled by the difference in
the respective $w(\vec r)$.

In order to obtain a Bose condensate excitonic insulator,
the temperature
must be sufficiently low and the lifetime of the excitons long enough.
Currently attainable experimental conditions
\cite{expt,expt1} satisfy the requirements: the
lifetime can be extended to $\sim 100~ns$, due to the
smaller electron-hole wavefunction overlap in the
double layers; and consequently thermal and chemical
equilibrium down to $1 K$ temperatures can be reached.
Even if some of the excitons may be trapped at the
impurity centers invariably present, a droplet of high density fluid should
still form, and be insensitive to the trapping.
One significant advantage of our proposal is that the
properties of the droplet, once formed, will be
relatively independent of the pump power, which
determines the average exciton density throughout the
sample, but does not affect the equilibrium chemical potential by much.
The excitonic, or the insulating nature, of the droplets could be
established by careful monitoring of evolution
of the luminescence spectrum edge with temperature \cite{ehRev}.
The Bose-Einstein condensation is a more
subtle and difficult issue to address due to
the reduced dimensionality
\cite{Thouless}. It may manifest itself in interesting ways
by its coupling to the photon field \cite{ZL}.

To summarize, we propose that a large droplet of a high density
excitonic insulator can be stablized
in a suitably arranged quantum
well structure where the electrons and the holes are spatially
separated. The argument given is quite general, and
the actual dimensions of the quantum well structures, while
affecting the density of the droplet, are not crucial
to its formation.
Such exciton droplets would be an example of the long sought-after
high density Bose condensed phase of excitons, and allow
one to examine the effects of coupling of a coherent
electron-hole system with light, which may be itself coherent,
as well as other interesting properties of the condensate.

\acknowledgements

We thank Dr. Phil Platzman
for useful discussions.
XJZ thanks Dr. Alfredo Pasquarello
and Dr. Stephen Fahy for encouragement
during the course of this work. Work at Rutgers
was supported by NSF through DMR92-21907.

\begin{figure}
\caption{Energy per electron-hole
pair ($E^{pair}$, solid lines)
and the chemical potential ($\mu$,
dashed lines)
of a double-layer electron-hole system as a function
of $r_s$ at separations $d = 0$ and $ d = 1.0$.
}
\label{Fig.1}
\end{figure}

\begin{figure}
\caption{
Gap function $\Delta_k$ and the pair-excitation energy
$E_k$ of the exciton droplet
as a function of $k$ for $r_s$ = 2.66 (dashed lines)
and for $r_s = 5.90$ (solid lines) at $d = 1.0$.
}
\label{Fig.2}
\end{figure}

\begin{figure}
\caption{
Schematic diagram of the $GaAs/AlGaAs$ quantum well structure
expected to contain the exciton
droplet at chemical equilibrium.
}
\label{Fig.3}
\end{figure}

\begin{references}
\bibitem{BiEx} We ignore the possible formation
of bi-excitons in this work.
It is disfavored by the
strong dipole-dipole repulsion in the
separate double-layer electron-hole system.
\bibitem{footnote} Theoretically, two parallel, metallic plates,
one with a Fermi surface of electrons and the other
of holes, were considered qualitatively in Ref.~\cite{Bilayer}
for possible superconductivity.
It was argued in Refs.~\cite{BOcond,JRK}
that there is no superconductivity in such systems.
\bibitem{factor2} These equations differ from the
usual BCS gap equations \cite{BCSbook}
by factors of two at several places, because
here we are explicitly calculating the energy of an
electron-hole pair,
rather than of a single electron participating in pairing.
\bibitem{spin} Not including the spin-degeneracy, the
densities are $1.3 \times 10^{11} / cm^2$ for
$r_s = 2.66$, and $2.6 \times 10^{10} /cm^2$
for $r_s = 5.90$ for $GaAs$ quantum wells.
\end{references}
\end{document}